\documentclass[chaos,twocolumn,amsmath,amsthm,amssymb,showpacs,floatfix]{revtex4-1}
\usepackage{graphicx}% Include figure files
\usepackage{dcolumn}% Align table columns on decimal point
\usepackage{bm}% bold math
\newcommand{\be}{\begin{equation}}
\newcommand{\ee}{\end{equation}}

\begin{document}

\title{Universal Fractional Map and Cascade of Bifurcations Type Attractors}

\author{M. Edelman}
%\email{edelman@cims.nyu.edu}

\affiliation{Department of Physics, Stern College at Yeshiva University, 245
  Lexington Ave, New York, NY 10016, USA 
\\ Courant Institute of
Mathematical Sciences, New York University, 251 Mercer St., New York, NY
10012, USA
}%

\date{\today}% It is always \today, today,

\begin{abstract}
We modified the way in which the Universal Map is obtained in the regular 
dynamics to derive the Universal $\alpha$-Family of Maps depending on a 
single parameter $\alpha > 0$ which is the order of the fractional
derivative in the nonlinear fractional differential equation describing 
a system experiencing periodic kicks. 
%Integer $\alpha>1$  implementations 
%of the Universal $\alpha$-Family of Maps are area preserving. 
We consider two particular $\alpha$-families
corresponding to the Standard and Logistic Maps. 
%In the case of integer $\alpha$ fractional Standard Map converges to the 
%Circle Map ($\alpha=1$), regular ($\alpha=2$) or three-  and higher-dimensional 
%($\alpha \ge 3$) Standard Maps, while the fractional Logistic Map converges 
%to the regular Logistic Map ($\alpha=1$) or  two- ($\alpha=2$)  
%and higher- ($\alpha \ge 3$) dimensional quadratic maps. 
For fractional $\alpha<2$
%$0<\alpha<1$ and $1<\alpha<2$ 
in the area of parameter values of the transition through the period 
doubling cascade of
bifurcations from regular to chaotic motion in regular dynamics
corresponding fractional systems demonstrate a new type of attractors -
cascade of bifurcations type trajectories.     
%\begin{description}
%05.45.-a,  05.45.Pq, 87.23.Cc
%\end{description}
\end{abstract}

%\pacs{05.45.-a,  05.45.Pq, 87.23.Cc}% PACS, the Physics and Astronomy
\maketitle

%\tableofcontents

{\bf  
Fractional dynamical systems (FDS) are
systems that can be described by  fractional differential equations (FDE)
with a fractional time derivative.
FDE are integro-differential equations and  solutions 
of the nonlinear FDE require long runs of computations.
This is why an investigation of the discrete maps which can be derived  
from the FDE, the fractional maps (FM), is even more important 
for the study of the general properties of the nonlinear FDS than
the investigation of the regular maps in the case of 
the regular nonlinear dynamical systems. 
In this article we investigate the  Universal $\alpha$-Family of Maps which
depends on a single parameter - the order $\alpha$ ($\alpha >0$) of 
the corresponding FDE with the periodic kicks.
We show that the integer members of the family represent
area/volume preserving maps and investigate their fixed/periodic
points. Using the particular examples of the Logistic and Standard 
$\alpha$-Families of Maps (S$\alpha$FM and L$\alpha$FM) 
we show how the maps' properties evolve with the increase in $\alpha$. 
The FDS are systems with memory and solutions of the FDE may possess quite
unusual properties: trajectories may intersect, attractors may overlap,
attractors exist in the asymptotic sense and their limiting values
may not belong to their basins of attraction.
Cascade of bifurcations type trajectories (CBTT) - are a new
type of attractors which exists only in the FDS. In a CBTT a cascade of
bifurcations occurs not as a result of a change in a system's parameter
(as in regular dynamical systems) but on a single attracting
trajectory during its time evolution. 
We show that the CBTT exist in both families for  
$0 < \alpha < 1$. When  $1 < \alpha < 2$ we found the
areas of parameters in which the  CBTT may exist in the S$\alpha$FM  
and the inverse CBTT in the L$\alpha$FM.  
The particular areas of the application of the FM may include
biological systems (population biology, human memory, adaptation) and 
fractional control. 
}

%%%%%%%%%%% Introduction %%%%%%%%%%%%%%%%%%%%%%%%%%%%%%%%

\section{Introduction}
Fractional derivatives (FD) are integro-differential operators 
in which an integral is a convolution of a function (or its derivative)
with a power function of a variable \cite{SKM,Podlubny,KST}.  
This is why fractional differential equations (FDE) are frequently used in 
science and engineering to describe systems with power law memory 
(see e.g. \cite{Podlubny,KST,Zbook,Hilfer,Advances,TarBook,Uch,Petras,PanDas}). 
We'll call systems which can be described by the FDE with a time FD fractional
dynamical systems (FDS). 
Because FDE are integro-differential equations and there are no
high order numerical algorithms to simulate such equations, derivation of
the fractional maps (FM) is important for the investigation of the general 
properties of the nonlinear FDS. The nonlinear FM are also discrete 
convolutions. They model systems in which the present state depends on a function
of all previous states weighted by a power of the time passed.
Systems with power law memory include viscoelastic materials 
\cite{MainardiBook2010}, electromagnetic fields in dielectric media
\cite{TDia2008a,TDia2008b,TDia2009}, Hamiltonian systems \cite{Zbook},
etc. 

There are many examples of systems with power law memory in biology. 
It has been shown recently 
\cite{Lund1,Lund2} that processing of external stimuli by individual neurons
can be described by fractional differentiation. 
There are multiple examples where power-law adaptation has been applied 
in describing the dynamics of biological systems at levels ranging from 
single ion channels up to human psychophysics
\cite{Wixted,Toib,Fairhall,Leopold,Ulanovsky,Zilany}. Fluctuations within 
single protein molecules demonstrate power-law memory kernel with the
exponent $-0.51 \pm 0.07$ \cite{Min}. The power law has been
demonstrated in many cases in the research on human memory. 
Forgetting - the accuracy in a memory task at time $t$ 
is given by $x=at^{-b}$, where $0<b<1$ 
\cite{Wixted,Wixted1,Wixted2,Rubin,Kahana}. Learning also can be
described by a power law. The reduction in reaction times that comes with
practice is a power function of the number of training trials \cite{Anderson}.

In many cases 
\cite{KST,KBT,TarMap} FDE are equivalent 
to the Volterra integral equations
of the second kind. This kind of equation (not necessarily FDE) is used in
nonlinear viscoelasticity (see for example \cite{Wineman2007,Wineman2009})
and in population biology and epidemiology see
\cite{HoppBook1975,PopBioBook2001}.  The very basic model in population
biology is  the ubiquitous Logistic Map. This map has been used to
investigate the essential property of the nonlinear systems - 
transition from order to chaos through a sequence of period-doubling
bifurcations, which is called cascade of bifurcations, and its relation to
the scaling properties of the corresponding systems (see
\cite{Universality}). But the subjects of population biology are always
systems with memory which can be related to changes in DNA or, as in the
case of human society, to legal regulations; and in most cases
reproduction also involves time delay. Development and investigation of a map
which would correspond to the Logistic Map with the power law memory and 
time delay
is important not only for the population biology but, as in the case of
regular dynamics, it is important in order to study the general properties
of the nonlinear FDS. One of the current main areas of the application 
of the nonlinear FDE,
control theory (see \cite{Petras,ControlBook2010}), will also
benefit from the study of the general properties of the FDS.

Nonlinear circuit elements with memory, memristors, memcapacitors, 
and meminductors \cite{Chua,VPC} can be used to model nonlinear systems with
memory. These elements may be common at the nanoscale, where the dynamical
properties of charged particles depend on the history of a system 
\cite{VPC}. Properties of such systems and their fractional
generalizations \cite{Machado,CafGra}
are already a subject of research but at present mathematical modeling
of the FM remains the most useful for the study of the general properties
of the FDS.

The first FM were derived from
the FDE 
%describing periodically kicked systems 
in \cite{TZ1,ETFSM,TEdisFSM,TarMap}. 
The first results of the investigation of
the FM (see \cite{ETFSM,TEdisFSM,myFSM,Taieb}) revealed new 
properties of the FDS: intersection of trajectories, overlapping of 
chaotic attractors,
existence of the attractors in the asymptotic sense (the limiting values
may not belong to their basins of attraction). 
Cascade of bifurcations type trajectories 
(CBTT) are the most unusual features of the investigated
FM. 
In the CBTT a cascade of bifurcations is not 
a result of the change in a system parameter (as in the regular dynamics) 
but appears as the attracting 
single trajectory and is a new type of attractors. 
All previous
investigations of the FM were done on the various forms of
the fractional two-dimensional Standard Map corresponding to the order
$1< \alpha \le 2$ of the fractional derivative.
%, including dissipative Standard Map. 
The CBTT appeared in all investigated FM. 
%and the major 
%conjecture made is that they appear in the area of the map parameters
%for which in the corresponding regular 2-D map a period doubling cascade 
%of bifurcations leads to the disappearance of a system of islands.
%The role of the cascades of bifurcations in the transition from order to
%chaos in the regular dynamics and their connection to the scaling
%properties of the corresponding systems are well investigated 
%(see \cite{Universality}). 
The consideration of the origin and the necessary and 
sufficient conditions of the CBTT's existence requires further investigation
of the FM, which includes development of the simple, if possible 
one-dimensional, FM. 
%The best investigated one-dimensional regular map is the ubiquitous Logistic
%Map. 
The Logistic Map, and the maps with $\alpha \le 1$ in general, can't be derived 
in a way previously used in \cite{TarBook,TarMap} to derive the FM for  
$\alpha > 1$ (for a detailed discussion see \cite{DNC}). 
In  \cite{DNC} we introduced the notions of the Universal Fractional Map of
an arbitrary order  $\alpha > 0$ and the $\alpha$-families of maps 
which allow a uniform derivation of the FM of the order $\alpha > 0$. 
In this paper we continue the investigation of the Universal Fractional
Map (Sec.~\ref{Universal}) and investigate the general properties (fixed and periodic points and their
stability) for the Universal Fractional Map of an arbitrary integer order
(Sec.~\ref{UniversalInteger}).  
We also conduct the detailed investigation of the members of the Logistic  
$\alpha$-Family of Maps (L$\alpha$FM) with $\alpha \le 2$ 
(Secs.~\ref{IntStLog}~and~\ref{CBTTsection}). As it has
been shown before for the members of the Standard 
$\alpha$-Family of Maps (S$\alpha$FM) with $\alpha \le 2$, in the L$\alpha$FM 
the CBTT exist for the fractional values of $\alpha$ but when $1< \alpha < 2$  
the L$\alpha$FM demonstrate only the inverse CBTT (Sec.{\ref{Logistic}).

\section{Universal Fractional Map}
\label{Universal}

To derive the equations of the Universal $\alpha$-Family of Maps (U$\alpha$FM) 
let's start with the equation introduced in \cite{DNC}:  
%differential 
\be
\frac{d^{\alpha}x}{dt^{\alpha}}+G_K(x(t- \Delta T)) \sum^{\infty}_{n=-\infty} \delta \Bigl(\frac{t}{T}-(n+\varepsilon)
\Bigr)=0,   
\label{UM1D2Ddif}
\ee
where $\varepsilon > \Delta > 0$,  $\alpha \in \mathbb{R}$, $\alpha>0$, in
the limit $\varepsilon  \rightarrow 0$. The initial conditions
should correspond to the type of fractional derivative we are going to
use. 
%\begin{figure}[!t]
%\includegraphics[angle=270, width=0.47\textwidth]{Jinan2.eps}
%%\includegraphics[width=0.47\textwidth]{Jinan1.eps}
%\vspace{-0.25cm}
%\caption{Here should be a caption}
%\label{fig:WE}
%\end{figure}
In the case $\alpha =2$, $\Delta = 0$, and $G_K(x)=KG(x)$ Eq.~\eqref{UM1D2Ddif} 
corresponds to the equation whose integration produces the regular
Universal Map (see \cite{Zbook}). Case $\Delta = 0$ and $G_K(x)=KG(x)$ has been
used 
%by Tarasov 
to derive  the fractional Universal Map for $\alpha >1$ (see Ch.~18 from
\cite{TarBook}). 
$\Delta \ne 0$ is
essential for the case $\alpha \le 1$ when $x(t)$ is a function
discontinued at the
time of the kicks \cite{TZ1,DNC} and the use of the
$K$ as a parameter rather than a factor is necessary to extend the class
of the considered maps to include the Logistic Map.   Without losing the
generality we assume $T=1$. Case $T \ne 1$ is considered in \cite{DNC} and 
can be reduced to this case by rescaling the time variable.
Further in the paper $T$ denotes periods of trajectories. 

%%%%%%%%%%%%%%%%%%%%%%% R-L %%%%%%%%%%%%%%%%%%%%%%%%%%%%%%%

\subsection{Riemann-Liouville Universal Fractional Map}
\label{UniversalRL}

In the case of the Riemann-Liouville fractional derivative  
Eq.~\eqref{UM1D2Ddif} can be written as   
\be
_0D^{\alpha}_tx(t) +G_K(x(t- \Delta )) \sum^{\infty}_{n=-\infty} \delta \Bigl(t-(n+\varepsilon)
\Bigr)=0,    
%, \    \ t \ge - \Delta T
\label{UM1D2DdifRL}
\ee
 where $\varepsilon > \Delta > 0$, $\varepsilon  \rightarrow 0$, $0 \le
 N-1 < \alpha \le N$, $\alpha \in \mathbb{R}$, $N \in \mathbb{Z}$, 
and the initial conditions $(_0D^{\alpha-k}_tx)(0+)=c_k$, $k=1,...,N.$
%\be
%(_0D^{\alpha-k}_tx)(0+)=c_k,  \    \ k=1,...,N.
%\label{UM1D2DdifRLic}
%\ee
The left-sided Riemann-Liouville fractional  derivative $_0D^{\alpha}_t x(t)$ defined for
$t>0$ \cite{Podlubny,SKM,KST} as 
{\setlength\arraycolsep{0.5pt}
\begin{eqnarray}
&&_0D^{\alpha}_t x(t)=D^n_t \ _0I^{n-\alpha}_t x(t) \nonumber \\
&&=\frac{1}{\Gamma(n-\alpha)} \frac{d^n}{dt^n} \int^{t}_0 
\frac{x(\tau) d \tau}{(t-\tau)^{\alpha-n+1}}~,
\label{RL}
\end{eqnarray}
}
where $n-1 \le \alpha < n$, $n \in \mathbb{Z}$,  
$D^n_t=d^n/dt^n$, $ _0I^{\alpha}_t$ is a fractional integral,
and $\Gamma()$ is the gamma function.

This problem 
%for a wide class of functions $G_K(x)$ 
can be reduced
\cite{TarBook,KST,KBT} to the Volterra integral equation of the second
kind for $t>0$
{\setlength\arraycolsep{0.5pt}
\begin{eqnarray}
&&x(t)= \sum^{N}_{k=1}\frac{c_k}{\Gamma(\alpha-k+1)}t^{\alpha -k} \nonumber \\ 
&&\hspace{-.2cm}-\frac{1}{\Gamma(\alpha)} \int^{t}_0 d \tau \frac{G_K(x( \tau - \Delta ))}{( t-\tau )^{1-\alpha}} \sum^{\infty}_{k=-\infty}
\delta \Bigl(\tau-(k+\varepsilon)\Bigr),
\label{VoltRL}
\end{eqnarray}
}
which integration gives ($t>0$)
{\setlength\arraycolsep{0.5pt}
\begin{eqnarray}
&&x(t)= \sum^{N-1}_{k=1}\frac{c_k}{\Gamma(\alpha-k+1)}t^{\alpha -k}  \nonumber \\ 
&&-\frac{1}{\Gamma(\alpha)}  
\sum^{[t-\varepsilon]}_{k=0} \frac{G_K(x (k+\varepsilon-\Delta))}{( t-(k+\varepsilon))^{1-\alpha}} 
\Theta(t-(k+\varepsilon)),
\label{VoltRLeq}
\end{eqnarray}
}
where $\Theta(t)$ is the Heaviside step function. In Eq.~\eqref{VoltRLeq}
we took into account that boundedness of $x(t)$ at $t=0$ requires $c_N=0$ and
$x(0)=0$. 

%With the introduction \cite{TZ1}
%\be
%p(t)= {_0D^{\alpha-N+1}_t}x(t),\ \ p^{(s)}(t)= {D^{s}_t}p(t),\ \ s=0,1,...,N-2 
%\label{FrMom}
%\ee 
%and 
%\be
%p^{(s)}(t)= {D^{s}_t}p(t),  \    \ s=0,1,...,N-2  
%\label{FrMoms}
%\ee 

With the introduction \cite{TZ1}
$p(t)= {_0D^{\alpha-N+1}_t}x(t)$, $p^{(s)}(t)= {D^{s}_t}p(t)$, $s=0,1,...,N-2$ 
%\label{FrMom}
%\ee 
%and 
%\be
%p^{(s)}(t)= {D^{s}_t}p(t),  \    \ s=0,1,...,N-2  
%\label{FrMoms}
%\ee 
Eq.~\eqref{VoltRLeq} leads to
{\setlength\arraycolsep{0.5pt}
\begin{eqnarray}
&&p^{(s)}(t)= \sum^{N-s-1}_{k=1}\frac{c_k}{(N-s-1-k)!}t^{N -s-1-k} \nonumber \\
&&\hspace{-0.3cm}-\frac{1}{(N-s-2)!}  
\sum^{[t-\varepsilon]}_{k=0} G_K(x(k+\varepsilon-\Delta))( t-k )^{N-s-2}, 
\label{VoltRLeqp}
\end{eqnarray}
}
%$$p^{(s)}(t)= \sum^{N-s-1}_{k=1}\frac{c_k}{(N-s-1-k)!}t^{N -s-1-k} $$
%\be
%-\frac{1}{(N-s-2)!}  
%\sum^{[t-\varepsilon]}_{k=0} G_K(x(k+\varepsilon-\Delta))( t-k )^{N-s-2}, 
%\label{VoltRLeqp}
%\ee
where $s=0,1,...,N-2$.
With the definitions $x_{n}=x(n)$ and $p^{(s)}_{n}=p^{(s)}(n)$
%\be
%x_{n}=x(n), \   \ p^{(s)}_{n}=p^{(s)}(n)
%\label{xnpn}
%\ee 
Eqs.~\eqref{VoltRLeq}~and~\eqref{VoltRLeqp} in the limit $\varepsilon  \rightarrow 0$ 
give for t=n+1 the Riemann-Liouville U$\alpha$FM (U$\alpha$RLFM) 
%equations
{\setlength\arraycolsep{0.5pt}
\begin{eqnarray}
&&x_{n+1}=  \sum^{N-1}_{k=1}\frac{c_k}{\Gamma(\alpha-k+1)}(n+1)^{\alpha -k} \nonumber \\  
&&-\frac{1}{\Gamma(\alpha)}\sum^{n}_{k=0} G_K(x_k) (n-k+1)^{\alpha-1}, 
\label{FrRLMapx} \\
&&p^s_{n+1}= \sum^{N-s-1}_{k=1}\frac{c_k}{(N-s-1-k)!} (n+1)^{N-s -1-k}
\nonumber \\  
&&-\frac{1}{(N-s-2)!}\sum^{n}_{k=0} G_K(x_k) (n-k+1)^{N-s-2}.
\label{FrRLMapp} 
\end{eqnarray} }
The map equations for momentum defined in a usual way
\be
p(t)= D^1_tx(t), \   \ p^s(t)= D^s_tp(t),  \    \ s=0,1,...,N-2,
\label{FrMomU}
\ee 
and the discussion on the different ways of the defining momentum in
the case of the Riemann-Liouville maps can be found in \cite{DNC}.
U$\alpha$RLFM Eqs.~\eqref{FrRLMapx}~and~\eqref{FrRLMapp} can be written in the 
much simpler form  
{\setlength\arraycolsep{0.5pt}
\begin{eqnarray}
&&p^s_{n+1}= p^s_n + \sum^{N-s-3}_{k=0}\frac{p^{k+s+1}_n}{(k+1)!} 
-\frac{G_K(x_n)}{(N-s-2)!}, 
\label{FrRLMappConv} \\
&&x_{n+1}=  \sum^{N-1}_{k=2}\frac{c_k}{\Gamma(\alpha-k+1)}(n+1)^{\alpha -k} \nonumber \\  
&&+\frac{1}{\Gamma(\alpha)}p^{N-2}_{n+1}+ \frac{1}{\Gamma(\alpha)} \sum^{n-1}_{k=0} p^{N-2}_{k+1}V^1_{\alpha}(n-k+1), 
\label{FrRLMapxConv} 
\end{eqnarray} 
}
where $s=0,1,...N-2$ and $V^k_{\alpha}(m)=m^{\alpha -k}-(m-1)^{\alpha -k}$. 
%\begin{equation} \label{V1}
%V^k_{\alpha}(m)=m^{\alpha -k}-(m-1)^{\alpha -k}. 
%\end{equation}

%%%%%%%%%%%%%%%%%%%%%%%% Integer R-L %%%%%%%%%%%%%%%%%%%%%%%%%%

\subsection{Integer-Dimensional Universal Maps}
\label{UniversalInteger}

For the integer 
%values of 
$\alpha=N$ the U$\alpha$FM
%Universal Map 
converges to   
{\setlength\arraycolsep{0.5pt}
\begin{eqnarray}
&&p^s_{n+1}= p^s_n + \sum^{N-s-3}_{k=0}\frac{ p^{k+s+1}_n}{(k+1)!} 
-\frac{G_K(x_n)}{(N-s-2)!}; 
\label{IntRLMappConv} \\
&&x_{n+1}=   x_n + \sum^{N-2}_{k=0}\frac{p^{k}_n}{(k+1)!} 
-\frac{G_K(x_n)}{(N-1)!} 
\label{IntRLMapxConv} 
\end{eqnarray} 
}
with the Jacobian ($N \times N$, $N \ge 2$) 
\[ \left| \begin{array}{ccccccccc}
1-\frac{\dot{G}_K(x)}{\Gamma(N)} & 1 & \frac{1}{2}  & ... & \frac{1}{\Gamma(n)} &
... & \frac{1}{\Gamma(N-1)} & \frac{1}{\Gamma(N)} \\
-\frac{\dot{G}_K(x)}{\Gamma(N-1)} & 1 & 1 & ... & \frac{1}{\Gamma(n-1)} &
... & \frac{1}{\Gamma(N-2)} & \frac{1}{\Gamma(N-1)} \\
-\frac{\dot{G}_K(x)}{\Gamma(N-2)} & 0  & 1 & ... & \frac{1}{\Gamma(n-2)} &
... & \frac{1}{\Gamma(N-3)} & \frac{1}{\Gamma(N-2)} \\
... & ... & ... & ... & ... &
... & ... & ... \\
-\frac{\dot{G}_K(x)}{\Gamma(N-k+1)} & 0 & 0 & ... & \frac{1}{\Gamma(n-k+1)} &
... & \frac{1}{\Gamma(N-k)} & \frac{1}{\Gamma(N-k+1)} \\
... & ... & ... & ... & ... & 
... & ... & ... \\
-\dot{G}_K(x) & 0 & 0 & ... & 0 &
... & 0 & 1 \end{array} \right| , \]
where $n$ and $k$
%$1<n<N$ and $1<k<N$ 
are the column and row numbers.
The first column can be written as the sum of the column with one in the
first row and the remaining zeros and the column which is equal to
$\dot{G}_K(x)$ times the last column. Determinants of the corresponding
matrices are 1 and 0; this is why the Jacobian is equal to one and the map
is the N-dimensional volume preserving map.

The integer U$\alpha$FM's fixed points are $p^s_0=0$ $(s=0,...,N-2)$ and 
$x_0$ satisfies $G(x_0)=0$. Their stability for $N \ge 1$ is defined by the
eigenvalues $\lambda$ of the Jacobian matrix $J(x_0,p^0_0,...,p^{N-2}_0)$. Polynomial
$P(\lambda)=\det[J(x_0,p^0_0,...,p^{N-2}_0)-\lambda I$] has values 
$P(0)=\lambda_1 \times ... \times \lambda_N =1$
%(volume preservation) 
and $P(1)=(-1)^N \dot{G}_K(x_0)$, which means that 
for odd values of $N>1$ stability is possible only if $\dot{G}_K(x_0)=0$.
For $T=2$ points $p^s_{n+1}=-p^s_n$ $(s=0,...,N-2)$ and  
$G(x_{n+1})=-G(x_{n})$. In the 
%particular 
case $N=3$ the only $T=2$
points are the fixed points.

%%%%%%%%%%%%%%%%% Caputo %%%%%%%%%%%%%%%%%%%%%%%%%%%%%

\subsection{Caputo Universal Fractional Map}
\label{UniversalC}

For Eq.~\eqref{UM1D2Ddif} with the left-sided Caputo derivative
\cite{KST} 
%\cite{Podlubny,SKM,KST} 
{\setlength\arraycolsep{0.5pt}
\begin{eqnarray}
&&_0^CD^{\alpha}_t x(t)=_0I^{n-\alpha}_t \ D^n_t x(t) \nonumber \\  
&&=\frac{1}{\Gamma(n-\alpha)}  \int^{t}_0 
\frac{ D^n_{\tau}x(\tau) d \tau}{(t-\tau)^{\alpha-n+1}}  \quad (n-1 <\alpha \le n)
\label{Cap}
\end{eqnarray}
}
the initial conditions may be taken as $(D^{k}_tx)(0+)=b_k$,  
$k=0,...,N-1$.
%\be
%(D^{k}_tx)(0+)=b_k,  \    \ k=0,...,N-1.
%\label{UM1D2DdifCic}
%\ee
This problem is equivalent to the Volterra integral equation of the second
kind ($t>0$)
{\setlength\arraycolsep{0.5pt}
\begin{eqnarray}
&&x(t)= \sum^{N-1}_{k=0}\frac{b_k}{k!}t^{k} \nonumber \\ 
&&\hspace{-0.4cm}-\frac{1}{\Gamma(\alpha)} \int^{t}_0 d \tau \frac{G_K(x( \tau - \Delta ))}{( t-\tau )^{1-\alpha}} \sum^{\infty}_{k=-\infty}
\delta \Bigl(\tau-(k+\varepsilon)\Bigr).
\label{VoltC}
\end{eqnarray}
}
With the introduction $x^{(s)}(t)=D^s_tx(t)$ the Caputo U$\alpha$FM 
(U$\alpha$CFM) 
%equations 
can be
derived in the form \cite{TarBook}
{\setlength\arraycolsep{0.5pt}
\begin{eqnarray}
&&x^{(s)}_{n+1}= \sum^{N-s-1}_{k=0}\frac{x^{(k+s)}_0}{k!}(n+1)^{k} \nonumber \\ 
&&-\frac{1}{\Gamma(\alpha-s)}\sum^{n}_{k=0} G_K(x_k) (n-k+1)^{\alpha-s-1},
\label{FrCMapx}
\end{eqnarray} 
}
where $s=0,1,...,N-1$.

\section{Integer-Dimensional Standard and Logistic Maps}
\label{IntStLog}

Fractional maps
Eqs.~\eqref{FrRLMappConv},~\eqref{FrRLMapxConv},~and~\eqref{FrCMapx}
are maps with memory in which the next value of the map variables depends
on all previous values. An increase in $\alpha$ leads to the increase in the
dimension of the map and to the increased power in the power law 
dependence of the weights of the 
old states (the increased role of memory). Integer values of $\alpha$
correspond to the degenerate cases in which map equations can be written
as the maps with full memory \cite{Memory} which are equivalent to the one
step memory maps in which map variables at each step accumulate 
information about all previous states of the corresponding systems 
(for a discussion on the fractional maps as 
maps with memory see \cite{DNC}). To fully understand the properties 
of the FM  we'll start with the consideration of the integer members of the
corresponding families of maps.

%Corresponding to the fact that 
%Because 
In the $\alpha=2$ case 
Eqs.~\eqref{IntRLMappConv}~and~\eqref{IntRLMapxConv} produce the 
Standard Map if $G_K(x)=K\sin(x)$ and in the $\alpha=1$ case the Logistic
Map results from $G_K(x)=x-Kx(1-x)$. We'll call the 
U$\alpha$FM Eqs.~\eqref{FrRLMappConv}~and~\eqref{FrRLMapxConv} 
with  $G_K(x)=K\sin(x)$
the Standard $\alpha$-RL-Family of Maps (S$\alpha$RLFM) and with  
$G_K(x)=x-Kx(1-x)$
the Logistic $\alpha$-RL-Family of Maps (L$\alpha$RLFM); 
we'll call  U$\alpha$FM Eq.~\eqref{FrCMapx} with  $G_K(x)=K\sin(x)$ 
the Standard $\alpha$-Caputo-Family of Maps (S$\alpha$CFM) and 
with  $G_K(x)=x-Kx(1-x)$
the Logistic $\alpha$-Caputo-Family of Maps (L$\alpha$CFM). 

For $\alpha=0$ the solution of Eq.~\eqref{UM1D2Ddif} is identical zero.
For $\alpha <1 $ the U$\alpha$RLFM  Eq.~\eqref{FrRLMapx}
also produces identical zero for maps which satisfy $G(0)=0$, which 
is true for the S$\alpha$RLFM and L$\alpha$RLFM.

There are no stable fixed points in the $\alpha=3$ Standard Map. 
%{\setlength\arraycolsep{0.5pt}
%\begin{eqnarray}
%&&x_{n+1}= x_n+p_{n+1}-\frac{1}{2}p^1_{n+1},\ \ ({\rm mod} \ 2\pi ),\nonumber\\
%&&p_{n+1}=p_n+p^1_{n+1},  \   \  ({\rm mod} \ 2\pi ), \\
%&&p^1_{n+1}=-K\sin(x_n)+p^1_n, \   \ ({\rm mod} \ 4\pi ).  \nonumber
%\label{3DSMn}
%\end{eqnarray}
%}
%For this map fixed points (including ballistic) are unstable.
For $K^2-16<4{p^1}^2<K^2$ there exist two lines of the stable $T=2$ on the 
torus ballistic points.
%{\setlength\arraycolsep{0.5pt}
%\begin{eqnarray}
%&&p^1_1,\  \ p_1=\frac{p^1_1}{2}-\pi(2n+1),\  \ K \sin x_1=2p^1_1,\\   
%&&p^1_2=-p^1_1, \  \ p_2=-\frac{p^1_1}{2}-\pi(2n+1),\  \ x_2=x_1-\pi(2n-1).\non%umber  
%\label{3DT2StablePoint}
%\end{eqnarray}
%}
For more on the preliminary results of the investigation of the 
S$\alpha$FM and L$\alpha$FM for $2<\alpha \le 3$ see \cite{DNC}.
A different form of the 3D Standard Map has been recently introduced 
and investigated in \cite{MeissSM} and some 3D quadratic 
volume preserving  maps were investigated in 
\cite{3DQ}.
The S$\alpha$FM and L$\alpha$FM with $\alpha>2$ are poorly investigated 
and 3D volume preserving maps in general are not fully investigated. 
In our simulations of the fractional maps we were able to find
the CBTT only for $\alpha  < 2$. This is why in the present article we 
won't further consider maps with  $\alpha  > 2$.

\subsection{One-Dimensional Maps}
\label{1D}

The  $\alpha=1$  S$\alpha$RLFM is a particular form of the Circle Map with 
zero driving phase 
\be
x_{n+1}= x_n - K \sin (x_n), \ \ \ \ ({\rm mod} \ 2\pi ).
\label{SM1D} 
\ee
\begin{figure}[!t]
\includegraphics[width=0.47\textwidth]{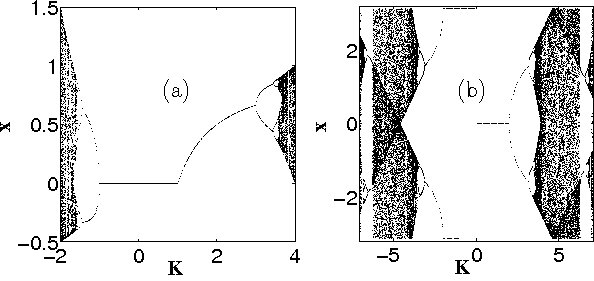}
\vspace{-0.25cm}
\caption{(a) The bifurcation diagram for the regular Logistic Map 
$x=Kx(1-x)$. 
(b) The bifurcation diagram for 1D Standard Map Eq.~(\ref{SM1D}). 
%(b) On $K-\alpha$ graph in S$\alpha$RLFM and S$\alpha$CFM CBTT  
%exist in the band of the map's parameters ending 
%at the cusp in the top right corner. Star marks the point 
%at which the Standard
%Map's ($\alpha=2$) $T=2$ elliptic points with $x_{n+1}=x_n-\pi$ and  
%$p_{n+1}=-p_n$ become unstable and bifurcate.   
}
\label{LS1D}
\end{figure}
The bifurcation diagrams for the regular Logistic Map and 
the one-dimensional Standard Map are presented in  Fig.~\ref{LS1D}.

The 1D Standard Map has the  attracting fixed points $2\pi  n$  
for $0<K <2$ and 
$ \pi + 2 \pi n$ when $-2 < K < 0$ (see Fig.~\ref{LS1D}b).
The antisymmetric $T=2$ points are stable for $2 < |K| < \pi$, while
$x_{n+1} = x_n+\pi$ sinks ($T=2$) are
stable when $\pi < |K| < \sqrt{\pi^2+2} \approx 3.445$. 
The stable $T=4$ sink appears at $K \approx 3.445$ and
the transition to chaos through the period doubling cascade of bifurcations
occurs at  $K \approx 3.532$. More on the properties of the $\alpha=1$  
Standard Map can be found in \cite{DNC}. 
. 

Stability properties of the Logistic Map are well known \cite{May}. For $K>0$, 
the $x=0$ fixed point is stable when  $K < 1$, 
the $(K-1)/K$ fixed point is stable when $1 < K < 3$,the  $T=2$ sink is stable
for $3 \le K < 1-\sqrt{6} \approx 3.449$, the $T=4$ sink is stable
for $3.449 < K < 3.544$, and
at $K  \approx 3.56995$ is the onset of
chaos, at the end of the period-doubling cascade of bifurcations.

\subsection{Two-Dimensional Maps}
\label{2D}

The regular ($\alpha=2$) Standard Map (Chirikov Standard Map) 
{\setlength\arraycolsep{0.5pt}
\begin{eqnarray}
&&p_{n+1}= p_{n} - K \sin x,  \ \ \ \ ({\rm mod} \ 2\pi ),  
\nonumber \\
&&x_{n+1}= x_{n}+ p_{n+1},  \ \ \ \ ({\rm mod} \ 2\pi )
\label{SMpx}
\end{eqnarray}}
%\be
%p_{n+1}= p_{n} - K \sin x,  \ \ \ \ ({\rm mod} \ 2\pi ) 
%\label{SMp}
%\ee
%\be
%x_{n+1}= x_{n}+ p_{n+1},  \ \ \ \ ({\rm mod} \ 2\pi ) 
%\label{SMx}
%\ee
demonstrates a universal generic behavior of the area-preserving maps whose
phase space is divided into elliptic islands of stability and areas of 
chaotic motion and is well investigated (see e.g. \cite{Chirikov}). 
In the S$\alpha$FM with $1 < \alpha < 2$ the elliptic islands
evolve into periodic sinks \cite{ETFSM,myFSM,Taieb,DNC}. The properties 
of the phase space and the appearance 
of the CBTT in the S$\alpha$FM are connected to the evolution (with 
the increase in parameter $K$) of the regular Standard Map's islands 
originating from the stable for $K<4$ fixed point (0,0). 
At $K=4$ it becomes unstable (elliptic-hyperbolic point transition) 
and two elliptic islands around the stable for $4 < K <2 \pi$  
period 2 antisymmetric ($p_{n+1}=-p_n$, $x_{n+1}=-x_n$) point appear. 
At $K=2 \pi$ this point turns into the  $T=2$ point 
with $p_{n+1}=-p_n$, $x_{n+1}=x_n-\pi$ which is stable for $2 \pi <K<6.59$.
The $T=4$ stable elliptic points appear at  $K \approx 6.59$ and the period
doubling cascade of bifurcations leads to the disappearance of
the islands of stability in the chaotic sea at $K \approx 6.6344$
\cite{Chirikov}. 
%\begin{figure}[!t]
%\includegraphics[width=0.47\textwidth]{Fig2.eps}
%\vspace{-0.25cm}
%\caption{ 
%Bifurcations in the 2D Logistic Map:
%(a) $T=4$ $\rightarrow$ $T=8$ bifurcation at $K \approx 5.527$; 
%(b )$T=8$ $\rightarrow$ $T=16$ bifurcation at $K \approx 5.5319$.   
%}
%\label{FigLog2D}
%\end{figure}

The $\alpha=2$   Logistic Map 
{\setlength\arraycolsep{0.5pt}
\begin{eqnarray}
&& p_{n+1}= p_n+Kx_n(1-x_n)-x_n, 
\nonumber \\
&& x_{n+1}= x_n + p_{n+1}
\label{LFMalp2}
\end{eqnarray}
}
is  a  quadratic area preserving map. The quadratic area
preserving maps with a stable fixed point at the origin  
were studied by H\'enon \cite{Henon69} and  a recent review
on quadratic maps can be found in \cite{ZeraS2010}. 
The map Eq.~\eqref{LFMalp2} has two fixed points: 
$(0,0)$ stable for $K \in (-3,1)$ and $((K-1)/K,0)$ stable for $K \in
(1,5)$. The $T=2$ elliptic point
{\setlength\arraycolsep{0.5pt}
\begin{eqnarray}
&&x = \frac{K+3 \pm \sqrt{(K+3)(K-5)}}{2K},  \nonumber \\ 
&&p=\pm \frac{\sqrt{(K+3)(K-5)}}{K}
\label{LFMalp2T2}
\end{eqnarray} 
}
is stable for $-2 \sqrt{5}+1<K<-3$ and $5<K<2 \sqrt{5}+1$.
The period doubling cascade of bifurcations (for $K>0$) with
further bifurcations,  $T=2$ $\rightarrow$ $T=4$ at $K \approx 5.472$,
$T=4$ $\rightarrow$ $T=8$ at $K \approx 5.527$, $T=8$ $\rightarrow$ $T=16$
at $K \approx 5.5319$, $T=16$ $\rightarrow$ $T=32$ at $K \approx 5.53253$,
etc., and the corresponding decrease in the area of the islands of stability
leads to chaos (see Fig.~\ref{FigLog2D}).
\begin{figure}[!t]
\includegraphics[width=0.47\textwidth]{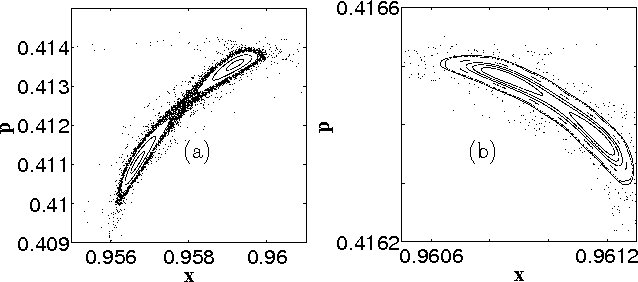}
\vspace{-0.25cm}
\caption{ 
Bifurcations in the 2D Logistic Map:
(a) $T=4$ $\rightarrow$ $T=8$ bifurcation at $K \approx 5.527$. 
(b) $T=8$ $\rightarrow$ $T=16$ bifurcation at $K \approx 5.5319$.   
}
\label{FigLog2D}
\end{figure}
%\be
%x = \frac{K+3 \pm \sqrt{(K+3)(K-5)}}{2K},  \ \ p=\pm \frac{\sqrt{(K+3)(K-5)}}{2%K}.
%\label{LFMalp2T2}
%\ee      
%For more on the $\alpha=2$ and $\alpha=3$ quadratic Logistic Maps see \cite{DNC}.
%{\setlength\arraycolsep{0.5pt}
%\begin{eqnarray}
%&&x_{n+1}=  x_n+y_{n+1}-\frac{1}{2}z_{n+1}, \ \ y_{n+1}=y_n+z_{n+1},
%\nonumber \\ 
%&&z_{n+1}=Kx_n(1-x_n)-x_n+z_n.
%\label{3DLMn}
%\end{eqnarray}
%}
%3D quadratic volume preserving  maps were investigated in 
%\cite{3DQ}.

\section{The Fractional ($\alpha<2$) S$\alpha$FM and L$\alpha$FM}
\label{CBTTsection}

\subsection{The CBTT in  the S$\alpha$FM and the L$\alpha$FM with $\alpha<1$}
\label{Alplt1}

\begin{figure}[!t]
\includegraphics[width=0.47\textwidth]{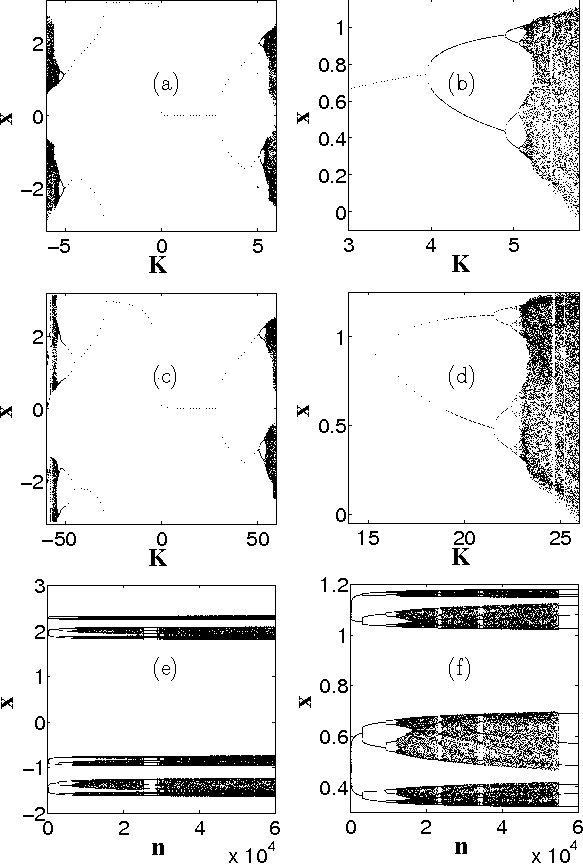}
\vspace{-0.25cm}
\caption{Bifurcations and the CBTT in the S$\alpha$CFM and the L$\alpha$CFM 
with $0<\alpha<1$. (a)-(d) bifurcation diagrams obtained after performing $10^4$
iterations on a single trajectory with $x_0=0.1$ for various values of $K$.  
(a) The S$\alpha$CFM with $\alpha=0.5$. 
(b) The L$\alpha$CFM with $\alpha=0.5$. 
(c) The S$\alpha$CFM with $\alpha=0.05$. 
(d) The L$\alpha$CFM with $\alpha=0.1$.  
(e) A CBTT in the S$\alpha$CFM with  $\alpha =0.01$ and $K=276$.
(f) A CBTT in  the L$\alpha$CFM with  $\alpha =0.1$ and $K=22.7$.   
}
\label{LowAlp}
\end{figure}
With the corresponding $G_K(x)$ the U$\alpha$CFM for $0<\alpha<1$ 
\be
x_{n+1}=  x_0- 
\frac{1}{\Gamma(\alpha)}\sum^{n}_{k=0} G(x_k) (n-k+1)^{\alpha-1}
\label{FrCMapxlt1}
\ee 
produces the S$\alpha$CFM 
\be
x_{n}=  x_0- 
\frac{K}{\Gamma(\alpha)}\sum^{n-1}_{k=0} \frac{\sin{x_k}}{(n-k)^{1-\alpha}},
 \   \  ({\rm mod} \ 2\pi )
\label{FrCSMlt1}
\ee 
and the L$\alpha$CFM
\be
x_{n}=  x_0+ 
\frac{1}{\Gamma(\alpha)}\sum^{n-1}_{k=0} \frac{Kx_k(1-x_k)-x_k}{(n-k)^{1-\alpha}},
\label{FrLMlt1}
\ee 
which are one dimensional maps
with the power law decreasing memory \cite{DNC}. 
The bifurcation diagrams for these 
maps are similar to the corresponding diagrams for the $\alpha=1$ case but are
stretched along the parameter $K$-axis and the stretchiness increases with 
the decrease in $\alpha$ Figs.~\ref{LowAlp}(a)-(d). In the area of the
parameter values for which on the bifurcation diagram stable periodic 
$T>2$ points exist individual trajectories are the CBTT  Figs.~\ref{LowAlp}(e),(f).

\subsection{The CBTT in the S$\alpha$FM  with $1< \alpha<2$}
\label{Standard}

The S$\alpha$RLFM and the S$\alpha$CFM with $1< \alpha<2$ were investigated in  
\cite{ETFSM,myFSM,Taieb}. In this subsection we'll recall some of the results of
this investigation. 
The fixed point $(0,0)$, which is a sink in this case, is stable for 
(see Fig.~\ref{alphaK}a)
\begin{equation} \label{Kc} 
0< K < K_{c1}(\alpha)= \frac{2 \Gamma(\alpha)}{V_{\alpha l}},
\end{equation}
where
\begin{equation} \label{Val} 
 V_{\alpha l}  =  \sum_{k=1}^{\infty} (-1)^{k+1} V_{\alpha}^1(k) \ \ .
\end{equation}
In accordance with  Sec.~\ref{IntStLog} $K_{c1}(1)=2$ and
$K_{c1}(2)=4$.
\begin{figure}[!t]
\includegraphics[width=0.47\textwidth]{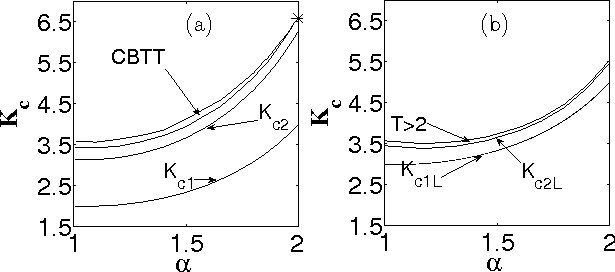}
\vspace{-0.25cm}
\caption{Bifurcations in the S$\alpha$FM and the L$\alpha$FM with
  $1<\alpha<2$. (a) The S$\alpha$FM  $K-\alpha$ graph.  The fixed point
  $(0,0)$ is stable for $K<K_{c1}$; the antisymmetric $T=2$ sink is stable
for   $ K_{c1}<K<K_{c2}$; two $T=2$ sinks $x_{n+1}=x_n-\pi$,
$p_{n+1}=-p_n$ are stable in a band above $K_{c2}$; the CBTT  
exist in the band of the map's parameters ending 
at the cusp in the top right corner; the upper curve is a border with chaos. 
The star marks the point ($K \approx 6.63$) at which the Standard
Map's ($\alpha=2$) $T=2$  points  become unstable and the
$T=4$ elliptic points are born.    
(b) The  L$\alpha$FM  $K-\alpha$ graph. One fixed point
is stable for $K<K_{c1L}$; the $T=2$ sink is stable for  
$K_{c1L}<K<K_{c2L}$; the sinks with $T \ge 4$ and the inverse CBTT 
exist in the upper band; the upper curve is a border with chaos. 
}
\label{alphaK}
\end{figure}
The antisymmetric period 2 sink
\begin{equation} \label{T2AS} 
p_{n+1} = -p_n, \    \  x_{n+1} = -x_n
\end{equation} 
is stable for $ K_{c1}(\alpha) < K < K_{c2}(\alpha)$ where  
$K_{c2}(\alpha)= 0.5 \pi K_{c1}(\alpha)$ 
with $K_{c2}(1)=\pi$ and $K_{c2}(2)=2\pi$.
\begin{equation} \label{T2nonAS} 
p_{n+1} = -p_n, \    \  x_{n+1} = x_n+\pi
\end{equation}
two T=2 sinks are stable in the band above
$K=K_{c2}(\alpha)$  curve (Fig.~\ref{alphaK}a). 
For $\alpha=1$
it corresponds to $\pi < |K| < \sqrt{\pi^2+2} \approx 3.445$ and
for the regular Standard Map the corresponding  elliptic points are stable 
when $2\pi < K < 6.59$. 

\begin{figure}[!t]
\includegraphics[width=0.47\textwidth]{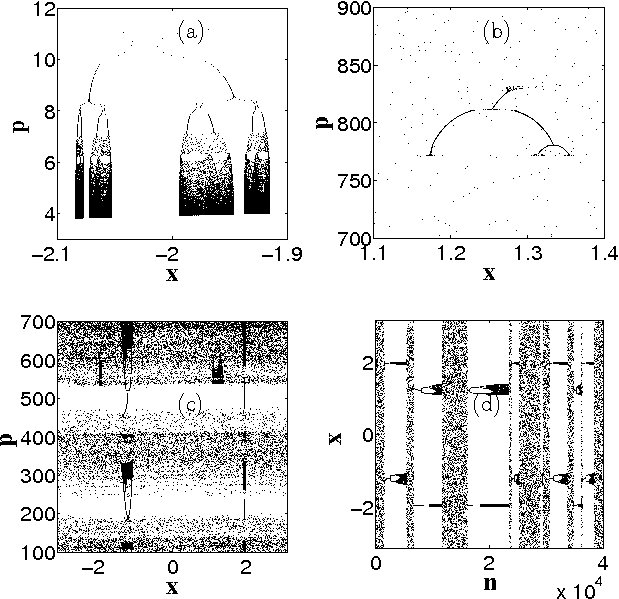}
\vspace{-0.25cm}
\caption{A single CBTT in the S$\alpha$RLFM. (a) One of the two branches of
the CBTT for  $\alpha=1.1, K=3.5$. 
(b) A zoom of a small feature in an intermittent 
trajectory for  $\alpha=1.95, K=6.2$.
(c) An intermittent trajectory in phase space for 
$\alpha=1.65, K=4.5$. 
(d) $x$ of $n$ for the case (c).   
}
\label{FSMcascades}
\end{figure}

For  $\alpha=1$ the $T=4$ sink appears at $K \approx 3.445$ and the transition
to chaos occurs at $K \approx 3.532$ (Sec.~\ref{1D}) while for  $\alpha=2$ the 
$T=4$ elliptic points appear at $K \approx 6.59$ and the sequence of
the period doubling 
bifurcations leads to the disappearance of the islands of stability in
chaotic sea at $K \approx 6.6344$ (Sec.~\ref{2D}).
For $1 < \alpha < 2$ the CBTT  exist in the band between two curves connecting
the above-mentioned points (Fig.~\ref{alphaK}a). 
Both curves are calculated numerically and confirmed 
by the large number of computer simulations \cite{myFSM,Taieb}. 
Within the CBTT band trajectories evolve from being very stable features 
which exist for the longest time we were running our codes, 500000
iterations, when $\alpha$ is close to 1 (Fig.~\ref{FSMcascades}a) to
being  barely distinguishable and short-lived features when $\alpha$ is
close to 2  (Fig.~\ref{FSMcascades}b). For the intermediate values of 
$\alpha$ CBTT behave similar to the sticky trajectories in Hamiltonian
dynamics:  occasionally 
trajectories enter CBTT and then leave them entering
the chaotic sea (Figs.~\ref{FSMcascades}c,~d).  

Let's list below some additional interesting properties of the  
S$\alpha$FM with $1<\alpha<2$ \cite{myFSM,Taieb}.
The types of solutions include periodic sinks, 
attracting slow diverging trajectories,
attracting accelerator mode trajectories, chaotic attractors, and the CBTT.
All attractors below the CBTT band are periodic sinks and slow diverging 
trajectories and all trajectories 
converge to one of those attractors. Each attractor has its own basin of
attraction and the chaotic areas exist in the sense that two trajectories
with infinitely close initial conditions from those areas may
converge to different attractors. Periodic sinks exist in the limiting sense
and the limiting values themselves in most of the cases do not belong to
their basins of attraction. The rate of convergence of trajectories to
the sinks depends on the initial conditions. The trajectories which start from
the basins of attraction converge fast as $\delta x \sim n^{-1-\alpha}$,
$\delta p \sim n^{-\alpha}$, while those starting from the chaotic areas
converge slow as  $\delta x \sim n^{-\alpha}$ 
(or even as $\delta x \sim n^{1-\alpha}$), $\delta p \sim n^{1-\alpha}$.
Trajectories may intersect and chaotic attractors overlap.
More on the 
properties of the  S$\alpha$RLFM and the S$\alpha$CFM for $1 \le \alpha \le 2$
can be found in \cite{myFSM,Taieb,DNC}.

\subsection{CBTT in  L$\alpha$FM with $1< \alpha<2$}
\label{Logistic}

In this part we'll investigate the L$\alpha$RLFM 
%\be \label{FLMRLp}
%p_{n+1} = p_n + K x_n (1-x_n)-x_n  \ \ ,
%\eqno (2)
%\ee
%\be
%\be \label{FLMRLx}
%x_{n+1} = \frac{1}{\Gamma (\alpha )} 
%\sum_{i=0}^{n} p_{i+1}V^1_{\alpha}(n-i+1) \ \ . 
%\frac{b}{\Gamma(\alpha-1)} (n+1)^{\alpha-2}.
%\eqno (3)
%\ee
{\setlength\arraycolsep{0.5pt}
\begin{eqnarray}
\label{FLMRLp}
&&p_{n+1} = p_n + K x_n (1-x_n)-x_n ,  \\
&&x_{n+1} = \frac{1}{\Gamma (\alpha )} 
\sum_{i=0}^{n} p_{i+1}V^1_{\alpha}(n-i+1).
\label{FLMRLx}
\end{eqnarray}
}
As in the case of the S$\alpha$FM, the partition of the phase space into
the areas of stability of the periodic sinks originating from the 
period one sink $(0,0)$ is almost the same (numerical result) 
for the L$\alpha$RLFM and the L$\alpha$CFM. For  $0<K<1$ 
all converging trajectories converge to $(0,0)$ point as 
$x \sim n^{-\alpha -1}$, $p \sim n^{-\alpha}$. For $1<K<K_{c1L}$ 
the only stable sink is the period one $((K-1)/K,0)$ sink and the rate of
convergence is  $\delta x \sim n^{-\alpha}$, $p \sim n^{-\alpha+1}$.
For   $K_{c1L}<K<K_{c2L}$ all converging  trajectories 
(this is a result from the large number of numerical simulations)
converge to the $T=2$
sink antisymmetric in $p$ Fig.~\ref{T2sink}a. 
\begin{figure}[!t]
\includegraphics[width=0.47\textwidth]{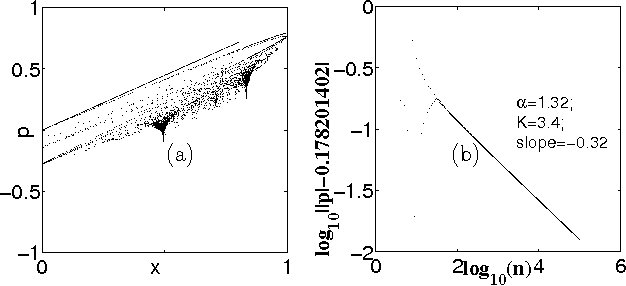}
\vspace{-0.25cm}
\caption{The L$\alpha$RLFM with $\alpha=1.32$, $K=3.4$.
(a) Phase space: 300 trajectories with $x_0=0$, $p_0=10^{-6}+0.00024i$, 
$0 \le i <300$. All converging trajectories converge to the $T=2$
antisymmetric in $p$ sink. 
(b) $\log p-\log n$ graph showing the rate of convergence $\delta p
\approx n^{-\alpha +1}$ on a single trajectory. 
}
\label{T2sink}
\end{figure}

To find the  L$\alpha$RLFM's critical curve $K_{c1L}$ on which, as a result of 
a bifurcation, the $T=1$ sink disappears and the $T=2$ sink is born, 
let's consider
the $T=2$ sinks. The results of large number of simulations
(see e.g. Fig.~\ref{T2sink}b) suggest the following asymptotic behavior:
\begin{equation} \label{P2Asy} 
p_n=p_l(-1)^n+ \frac{A}{n^{\alpha -1}} \ \  .
\end{equation}
Then, from   Eq.~(\ref{FLMRLx})
{\setlength\arraycolsep{0.5pt}
\begin{eqnarray}
&&x_{lo}=\lim_{n \rightarrow \infty}x_{2n+1}=\frac{p_l}{\Gamma(\alpha)} 
\lim_{n \rightarrow \infty}\sum^{2n+1}_{k=1}(-1)^k V_{\alpha}^1(k) \nonumber \\ 
&&\hspace{-0.2cm}+\frac{A}{\Gamma(\alpha)}\lim_{n \rightarrow \infty}
\sum^{2n-1}_{k=1}\frac{\alpha-1}{k^{\alpha-1}(2n-k)^{2-\alpha}}=
-\frac{p_l}{\Gamma(\alpha)}V_{\alpha l}  
\label{limxo} \\
&&\hspace{-0.2cm}+\frac{(\alpha-1)A}{\Gamma(\alpha)} \int^{1}_0
\frac{x^{1-\alpha} dx}{(1-x)^{2-\alpha}}
=-\frac{p_l}{\Gamma(\alpha)}V_{\alpha l}+A \Gamma(2-\alpha). \nonumber
\end{eqnarray}
}
In a similar way 
\begin{equation} \label{limxe} 
x_{le}=\lim_{n \rightarrow \infty}x_{2n}=\frac{p_l}{\Gamma(\alpha)}V_{\alpha l}+A \Gamma(2-\alpha).
\end{equation}
In the limit $n \rightarrow \infty$   Eq.~(\ref{FLMRLp}) gives
{\setlength\arraycolsep{0.5pt}
\begin{eqnarray}
\label{limpo}
&&-2p_{l}=Kx_{le}(1-x_{le})-x_{le} ,  \\
&&2p_{l}=Kx_{lo}(1-x_{lo})-x_{lo}.
\label{limpe}
\end{eqnarray}
}
The system of Eqs.~(\ref{limxo})-(\ref{limpe}) has four equations and four
unknown variables $p_{l}$, $A$, $x_{lo}$, and $x_{le}$.
This equation has two obvious solutions  $x_{lo}=x_{le}=p_{l}=A=0$ 
and $x_{lo}=x_{le}=x_l=(K-1)/K$, $p_{l}=0$, $A=x_l/\Gamma (2-\alpha)$,
corresponding to two fixed points. If  $x_{lo} \ne x_{le}$, then
\begin{equation} \label{A} 
A= \frac{K-1+\frac{2\Gamma(\alpha)}{V_{\alpha l}}}{2K\Gamma(2-\alpha)}
\end{equation}
and $x_{le}$ is a solution of the quadratic equation
{\setlength\arraycolsep{0.5pt}
\begin{eqnarray}
&&x_{le}^2-\Bigl(\frac{2\Gamma(\alpha)}{KV_{\alpha l}}+
\frac{K-1}{K}\Bigr)x_{le}+  \Bigl(\frac{\Gamma(\alpha)}{2KV_{\alpha l}}+
\frac{K-1}{4K}\Bigr)^2   \nonumber \\ 
&&\hspace{-0.2cm} -\frac{(K-1)\Gamma(\alpha)}{K^2V_{\alpha l}}-
\frac{(K-1)^2}{2K^2}=0,
\label{x}
\end{eqnarray}
}
which for positive $K$ has solutions only when 
\begin{equation} \label{KcL} 
K  \ge K_{c1l}=1+\frac{2\Gamma(\alpha)}{V_{\alpha l}}.
\end{equation}
Direct numeric simulations of the map Eqs.~(\ref{FLMRLp}) and (\ref{FLMRLx})
confirm this  $K_{c1l}$ value as well as the limiting values for $p_l$, 
$x_{lo}$, and $x_{le}$. For a way to calculate numerically 
slow converging series Eq.~(\ref{Val}) for $V_{\alpha l}$ see APPENDIX.

In the CBTT band of the L$\alpha$FM, the narrow band between the upper two
curves on Fig.~\ref{alphaK}b, the cascade of bifurcation type trajectories
exist only in the form of the inverse CBTT (see Fig.~\ref{inv}).
The inverse CBTT which exist for the L$\alpha$CFM 
(Figs.~\ref{inv}~and~\ref{CvsRLinv}a) are almost impossible to find in the  
L$\alpha$RLFM (Fig.~\ref{CvsRLinv}b). The closer $\alpha$ is to two the more
difficult it is to find the CBTT in the phase space or $x$-$n$ graph 
of the L$\alpha$CFM.
\begin{figure}[!t]
\includegraphics[width=0.47\textwidth]{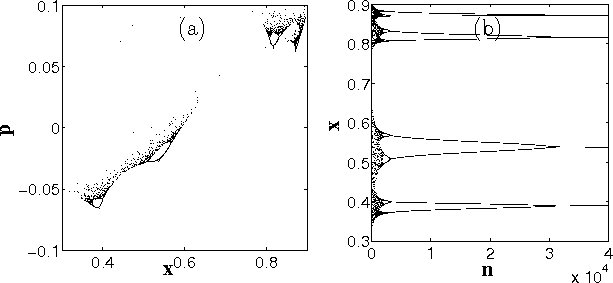}
\vspace{-0.25cm}
\caption{An inverse CBTT in the L$\alpha$CFM with $\alpha=1.2$, $K=3.45$.
40000 iterations on a trajectory with $x_0=0.01$ and $p_0=0.1$. (a) Phase
space. (b) $x-n$ graph. 
}
\label{inv}
\end{figure}
\begin{figure}[!t]
\includegraphics[width=0.47\textwidth]{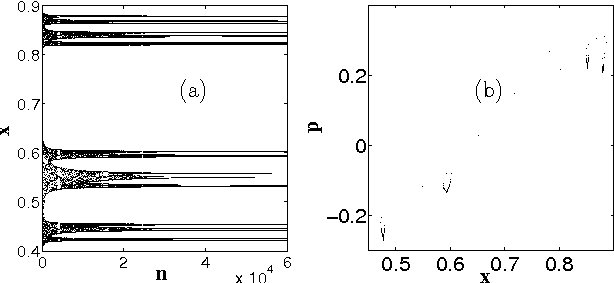}
\vspace{-0.25cm}
\caption{The L$\alpha$CFM vs. the L$\alpha$RLFM.  (a) 60000
iterations on a single L$\alpha$CFM trajectory for $\alpha=1.6$ $K=3.9$  
(b) the $T=4$ trajectory for the L$\alpha$RLFM with $\alpha=1.6$ $K=3.88$. 
}
\label{CvsRLinv}
\end{figure}

\section{Conclusion}
\label{Conclusion}

The Universal  $\alpha$-Family of Maps introduced in this paper is the 
extension of the fractional Universal Map, which allows consideration of the
Logistic Map as its particular form. The results of the investigation of 
the Standard and Logistic 
Families of Maps suggest that the existence of the cascade of bifurcations
type trajectories is a general property of the fractional dynamical systems. 
They appear for the parameter values
corresponding to the transition through the period doubling cascade of
bifurcations from regular to chaotic motion in the regular dynamics.
Fig.~\ref{LowAlp} and
Fig.~\ref{FSMcascades} support our statement that 
with the increase in $\alpha$, which represents the increase in the systems' 
dimension and memory (increase in the weights of the earlier states),
systems demonstrate more complex and chaotic behavior. 
Biological systems are systems with memory and the Fractional Logistic Map
can serve as a basic model in  population biology with memory.
We believe that experiments on human memory and/or adaptive biological systems, 
which in many respects 
are  systems with power law memory, could demonstrate the CBTT-like
behavior.
New types of materials with memory, such as  memristors, memcapacitors,
and meminductors, could be used to model fractional systems 
to demonstrate the existence of the CBTT. The
$\alpha > 2$ Standard and Logistic Maps (including their integer volume
preserving forms)
are topics of ongoing research and their further investigation 
is necessary to demonstrate the consistency of the changes in the
properties of the fractional systems with the change in $\alpha$.

\begin{acknowledgments}
The author expresses his gratitude to V.~E. Tarasov for the useful
remarks and to E. Hameiri and H. Weitzner 
for the opportunity to complete this work at the Courant Institute.
\end{acknowledgments}

\section*{Appendix}
\label{app}

$V_{\alpha l}$ can be written as
\be \label{1A} 
V_{\alpha l} =  \sum_{k=1}^{\infty} (-1)^{k+1} V_{\alpha}(k) = S_1+S_2,
%\eqno (1A)
\ee
where
\be \label{2A} 
S_1 =  \sum_{k=1}^{2N} (-1)^{k+1} V_{\alpha}(k) 
%\eqno (2A)
\ee
with the $N$ sufficiently large and 
\be \label{3A} 
S_2 =  \sum_{k=N+1}^{\infty} \{V_{\alpha}(2k-1)-V_{\alpha}(2k)\} \   \ .
%\eqno (3A)
\ee
The value of $S_1$ can be directly calculated numerically with high precision. 
The second sum can be developed into a series as follows
{\setlength\arraycolsep{0.5pt}
\begin{eqnarray}
&&S_2 = \sum_{k=N+1}^{\infty}
(2k)^{\alpha-3}(\alpha-1)(2-\alpha)\Bigl(1+\frac{3-\alpha}{2} \frac{1}{k}
\nonumber \\ 
&&+\frac{7(3-\alpha)(4-\alpha)}{48}\frac{1}{k^2}+  
\frac{(3-\alpha)(4-\alpha)(5-\alpha)}{32}\frac{1}{k^3}+O(\frac{1}{k^4})\Bigr)
\nonumber \\ 
&&=
(2)^{\alpha-3}(\alpha-1)(2-\alpha)\Bigl(\zeta(3-\alpha)+\frac{3-\alpha}{2}\zeta(4-\alpha)+ \label{4A} \\
&&\frac{7(3-\alpha)(4-\alpha)}{48}\zeta(5-\alpha)+ 
\frac{(3-\alpha)(4-\alpha)(5-\alpha)}{32}\zeta(6-\alpha)\Bigr) 
\nonumber \\
&& -\sum_{k=1}^{N}
(2k)^{\alpha-3}(\alpha-1)(2-\alpha)\Bigl(1+\frac{3-\alpha}{2} \frac{1}{k} 
+\frac{7(3-\alpha)(4-\alpha)}{48}\frac{1}{k^2}
\nonumber \\
&&+  \frac{(3-\alpha)(4-\alpha)(5-\alpha)}{32}\frac{1}{k^3}\Bigr)+O(\frac{1}{N^{6-\alpha}}).\nonumber
\end{eqnarray}
}  
This is what finally was coded using a fast method for
calculating values of the $\zeta$-function.

\end{document}